**Beyond the Human-AI Binaries:**

**Advanced L2 Writers' Self-Directed Use of Generative AI in Academic Writing**

Chaoran Wang, Colby College, chaowang@colby.edu

Wei Xu, Northern Illinois University, weixu@niu.edu

Xiao Tan, Utah State University, xiao.tan@usu.edu

**Abstract**

This study explores the self-directed use of Generative AI (GAI) in academic writing among advanced L2 English writers, challenging assumptions that GAI undermines meaningful learning and holds less value for experienced learners. Through case studies, we investigate how three (post)doctoral writers engage with GAI to address specific L2 writing challenges. The findings revealed a spectrum of approaches to GAI, ranging from prescriptive to dialogic uses, with participants positioning AI as a tool versus an interactive participant in their meaning-making process, reflecting different views of AI as a mechanical system, social construct, or distributed agency. We highlight the ways AI disrupts traditional notions of authorship, text, and learning, showing how a poststructuralist lens allows us to transcend human-AI, writing-technology, and learning-bypassing binaries in our existing discourses on AI. This shifting view allows us to deconstruct and reconstruct AI's multifaceted possibilities in L2 writers' literacy practices. We also call for more nuanced ethical considerations to avoid stigmatizing L2 writers' use of GAI and to foster writerly virtues that reposition our relationship with AI technology.

**1. Introduction**

In a postdigital world where a nexus of artificial intelligence, algorithms, and hybrid spaces are mediating writers' literacy experiences, new challenges and opportunities arise in academic writing. Generative AI (GAI) technologies have opened new avenues for L2 writers to seek assistance with easy access. Yet, gaps in understanding these writing practices persist, partly due to the largely self-directed nature in which GAI was used. A recent survey published in *Nature* highlights this trend, revealing that early-career scholars in social sciences and engineering often rely on self-taught, self-regulated approaches when using GAI in their work (Nordling, 2023). However, AI usage remains a highly controversial and gray area in high-stakes academic writing. While some writers worry that the disclosure of AI usage may carry potential risks, if not, at least ambiguous negative connotations regarding their academic rigor and integrity, others opt to use AI without disclosure (Conroy, 2023).

This situation highlights the urgency to examine the nuanced and contextualized GAI use in academic settings. Why do L2 scholarly writers turn to GAI, and how do they engage with this technology? We undertook the current study to explore advanced L2 writers' self-directed use of GAI by focusing on three (post)doctoral L2 English writers' experiences. By unpacking these experiences, we aim to offer insights into the role GAI plays in L2 academic writing and to rethink the evolving relationship of writing, technology, and learning.

**2. Literature Review**

*2.1 GAI in academic writing practices*

GAI is increasingly influencing academic writing practices and has become part of writers' evolving academic repertoire for competence and communication (e.g., Ou et al., 2024; Author, 2024). Optimists suggest that large language models (LLMs) could act as democratizing agents for L2 scholars in English-dominant academic writing, research, and publishing. Hwang et al. (2023) argued that LLMs could "lower the barrier to academic writing in English" (p. 952), while Giglio and Costa (2023) believed that AI has the potential for "facilitating composition by nonnative English speakers and decreasing inequalities existing with native-speaking researchers" (p. 4). In a review of 24 studies on AI's impact on academic writing and research, Khalifa and Albadawy (2024) position AI as a "productivity tool" that revolutionizes academic writing by supporting researchers in various dimensions of their work, including idea generation, content structuring, literature synthesis, data management and analysis, editing and publishing, and communication and outreach.

However, such involvement of GAI in writing introduces concerns regarding knowledge production and possible learning loss for L2 writers. Floridi (2023) described AI as "agency without intelligence," contending that "we have decoupled the ability to act successfully from the need to be intelligent, understand, reflect, and consider anything" (p. 5). This critique is particularly pertinent to educational institutions, where writing has long been central to both demonstrating and evaluating students' disciplinary learning and knowledge. Similar apprehensions appeared in scholarly writing, where writing and publishing also serve as a primary metric of one's intellectual contribution and scholarly authority. Academic genres, in this sense, have been functioning as an institution that affirms one's legitimate membership within the academic community by textualizing its established way of thinking and articulating. As scholars decried the "death" of essays as an effective way to assess learning and the potential of AI for students to "avoid education" (Gefen & Arinze, 2023), reviewers, editors, and other stakeholders in academic communities cautioned that involving GAI in academic knowledge creation and dissemination may risk low-quality writing, problematic research process, and even fake science (e.g., Kim et al., 2024).

These negative connotations around AI use make academic writers wary of openly sharing their AI-assisted writing experiences, as they worry it may undermine their credibility, ethics, and academic rigor (Conroy, 2023; Author, 2024). The hesitance is echoed in Warschauer et al.'s (2023) assertion that GAI amplifies the risk of writers, particularly L2 writers, being accused of plagiarism when incorporating AI-generated text–a bias further compounded by AI detection tools that often exhibit prejudices against L2 writers (Liang et al., 2023). These point to broader issues regarding fairness in the academic evaluation of AI-assisted writing and how a lack of understanding of AI-assisted L2 writing practices may reinforce existing stigmas about L2 writers (Flowerdew, 2008; Pecorari & Petrić, 2014).

While concerns about GAI's role in academic writing and knowledge production persist, another issue emerges regarding the potential impact of GAI on writing voice and linguistic diversity (Kuteeva & Andersson, 2024; Author, 2024). Academic writing has historically privileged Western epistemologies, Western rhetorics, and standardized linguistic norms. The integration of GAI and LLMs into the writing process may further reinforce this homogenization by reproducing and promoting the language patterns on which these models are trained—typically based on dominant, standardized forms of English and embedded with various cultural, racial, and linguistic biases. Scholars have criticized the "whiteness" of AI (Cave & Dihal, 2020) and urged us to consider how using GAI as part of academic writing practice could affect our language choices, leading to less diverse and more predictable linguistic patterns. Incorporating the AI language may rob L2 writers of their distinctive style and voice and the opportunities to look inward at the language diversity and creative expressions inherent in themselves. This poses threats to both knowledge creation and language diversity.

## 2.2 Self-directed learning

Self-directed learning (SDL) is a concept that has been widely researched over the past decades, grounded in the premise that everyone is fundamentally a lifelong learner. Digital technologies have become an integral part of our lives, mediating self-directed experiences of literacy and learning practices. Meanwhile, learning how to learn with evolving digital technologies per se—a process of developing digital literacy and skills—is also a crucial aspect of SDL.

Numerous research has suggested that self-direction is a critical quality in leveraging technology for personal growth and learning (Li & Bonk, 2023). Garrison's (1997) seminal work conceptualizes SDL as "an approach where learners are motivated to assume personal responsibility and collaborative control of cognitive (self-monitoring) and contextual (self-management) processes in constructing and confirming meaningful and worthwhile learning outcomes" (p. 18). Song and Hill (2007) noted that while Garrison's (1997) model emphasized SDL as both a personal attribute and a process involving resource management and learning strategies, it did not give sufficient attention to the influence of contextual factors. Underscoring the importance of context in SDL, Song and Hill reframed the model into three key constructs: (1) personal attribute such as motivation, resource and strategy use, and learners' prior knowledge and experience, (2) autonomous process which involves planning, monitoring, and evaluating, and (3) contextual factors that focus on the design elements, the learning context, and support elements (e.g. peers and instructors) that may impact learners' SDL.

Considering the growing adoption of GAI in language learning, Author (2024) expanded upon Song and Hill's (2024) model and proposed an AI-assisted SDL framework based on their research on language teachers and learners. Author (2024) identified key personal attributes that may influence a learner's self-directed use of AI, including their attitude toward AI technology, self-efficacy beliefs about their competence to use AI, motivation, and the use of resources and strategies. They modified the autonomous process into an autonomous-adaptive process, highlighting that working with GAI demands adaptability to continually adjust and refine one's use in response to AI's outputs as well as its fast-evolving nature and the potential advancement of its computational agency (Author, 2024). In the third component, contextual factors, they pointed out the key design elements that shaped GAI, such as training data, algorithms, and the AI interface. They also emphasized the complexity of social elements in AI-assisted SDL, which span multiple levels, such as institutional policy on AI adoption, social justice considerations, ethical considerations, and cultural and geopolitical norms.

Today's L2 writers are exposed to a variety of AI technologies outside formal educational settings and are increasingly engaged in self-directing their learning and literacy practices with these technologies. GAI's language affordances and its personalized, self-paced, and interactive nature offer unique opportunities for SDL. It also extends beyond a linguistic or editing tool and has been found to be involved in writers' various components of writing and research, such as sourcing information, summarizing literature, and preparing manuscripts (e.g., Nordling, 2023; Author, 2024). These practices blur the boundaries between human authorship and AI-generated content, and if used improperly, risk the potential encroachment on knowledge creation and the expression of an authentic writing voice. These complicated issues need to be examined in light of L2 writers' specific writing processes and practices given their writing contexts, purpose, and audiences.

We undertook the current study to explore advanced L2 writers' self-directed use of GAI by focusing on three doctoral and postdoctoral L2 English writers from different disciplines. While many studies have explored L2 students' use of AI in formal educational contexts, there has been less research on the self-directed GAI use in L2 scholars' academic work. In this study, we draw upon the concept of SDL, which allows us to consider the various factors that may shape L2 writers' AI-assisted academic writing given their personal experiences, goals, and contexts. In alignment with previous scholarships on AI, writing, and SDL, we ground our study in the social nature of AI-assisted writing (Author, 2024), meaning it goes beyond improving one's writing products and entails navigating evolving technologies as part of one's rhetorical and writing repertoire throughout one's lifelong learning journey. The central questions guiding our inquiry are: How did advanced L2 writers self-direct their use of GAI technology for academic writing? What factors shaped the L2 writers' self-directed use of GAI for academic writing? By exploring these questions, we aim to shed light on the role of GAI in L2 junior scholars' academic literacies and provide implications for L2 academic writing.

**3. Methods**

To trace participants' writing practices and provide a comprehensive understanding, we adopt a multiple-case study approach (Prior, 1995). A multiple case study design helps understand a phenomenon through highly contextualized illustrations of the experiences of multiple cases by in-depth drawing on more than one data source (Negretti & Khuder, 2023). Thus, it allows us to capture the complex and varied ways in which L2 writers engage with AI, providing a contextualized understanding of their practices within their own rhetorical contexts and disciplines.

### *3.1 Participants*

Our participants are three L2 Chinese writers currently studying or working in the United States. Due to the high-stakes nature of academic research writing, the sensitivity surrounding GAI use in academic work, and concerns about protecting the participants' identities, we chose to eliminate detailed information about the participants' backgrounds such as their institutions and specific research disciplines and topics. We chose the participants based on convenience and purposeful sampling, aiming to include L2 scholars from diverse disciplines with varying experience and backgrounds in academic English writing. All of the participants engaged with AI tools for a range of academic purposes and genres on a regular basis. Table 1 summarizes the participants' professional information, educational backgrounds, publication experiences, and their disciplinary contexts.

**Table 1.**

*A Summary of Participants' Professional Information, Education and Publication Experiences, and Disciplines*

| Name | Professional information | Educational Background | Publication Experiences | Discipline |
|------|------------------------|------------------------|-------------------------|------------|
| Mei | Assistant Professor at a private institution in the Eastern US | Received undergraduate and master's degrees in China<br><br>Received a PhD degree in the US | Published several articles in major journals in her field<br><br>Currently working on a book project developed from her dissertation study for her tenure case | Humanities |
| Daniel | PhD student at a large public research institution in the midwestern US | Completed undergraduate and master's degrees in the US<br><br>Taking the last course for his PhD program during the study. The course was outside his discipline but he wanted to explore it. | Had not yet published peer-reviewed articles, but had independently conducted research and written many research papers from his coursework | Social Sciences |
| Alex | Postdoctoral fellow working as a data scientist in a public sector in the western US | Received undergraduate and master's degrees in China<br><br>Recently graduated from his PhD program | Began to publish in his field before coming to the US, mostly in his L2<br><br>Prolific in English publications in peer-reviewed STEM journals | STEM |

### 3.2 Data collection and analysis

Over a four-month period from February 2024 to May 2024, we collected data including participants' AI chat history, their written drafts, and transcripts of process-tracing interviews (Prior, 2004). All three participants primarily used ChatGPT in academic writing, and they shared their AI chat history with us. Process-tracing interviews were used to capture the processes, strategies, and decisions that participants made while working on their writing projects. Data were collected through bi-weekly research interviews with each participant. During each interview, the participant shared: 1) any writing tasks they had worked on during the period, 2) any AI usage logs and written drafts they were willing to provide, and 3) descriptions of their AI use process, goals, feelings, experiences, and reflections on using GAI for academic writing. Each process-tracing interview varied in length, depending on how much writing the participant shared during the two-week period. The data collection yielded 156 pages of AI logs, 405 minutes of interview recordings, and 76 pages of written drafts.

The data analysis included both protocol and open coding. To answer the first and the second research questions, a coding scheme with elements from Garrison's (1997) and Author's (2024) SDL frameworks was developed and used to guide the protocol coding of the interview data. For the third research question, we conducted open coding of interview transcripts following Creswell and Poth's (2018) thematic analysis. After coding the interview data, we triangulated our analysis with other data sources to strengthen the findings. To minimize researcher bias and enhance the trustworthiness of the study, we conducted member checks with the participants to confirm the accuracy of the data and interpretations.

## 4. Findings

In this section, we unpack each participant's self-directed use of GAI in their writing practices, showing their specific approaches and the contextual factors shaping their AI usage for academic writing.

### 4.1 Mei

Mei's entering motivation for engaging with GAI was to address challenges in articulating her ideas in writing. She viewed AI as a helpful "stimulus" that allowed her to find the "right words to express myself." She typically began by organizing her thoughts in ways that felt close to what she wanted to express, describing it as "writing my string of thoughts into very poor English" that "only" she could understand. Then she would attempt her prompts in two ways: by asking ChatGPT if it understood her input and having it rephrase her ideas in its own terms, or by asking ChatGPT to "edit to enhance clarity." She evaluated the AI's outputs by gauging whether there were misunderstandings between her intended meaning and ChatGPT's interpretation. When sensing a discrepancy, she would continue the conversation, clarifying what she really meant. Mei acknowledged that evaluating ChatGPT's wording often inspired her to reflect on her own argument and language, asking herself: "Is this what I actually want to express?" "Why did AI use this word to represent what I just said?" Mei's AI usage centered around linguistic practices, but it went beyond one-way, mechanical editing for linguistic accuracy. Describing the process as "communicating with AI," Mei's approach features a heuristic of meaning co-construction with AI through fine-tuning its outputs to achieve mutual understanding.

Mei's self-directed use of AI also involved adapting to its limitations and adjusting her learning strategies accordingly. For instance, if AI altered her sentence to the point that it no longer "sounded human," she would revise her prompt, specifying, "edit it but keep my language as much as possible." Even when AI failed to capture her meaning and she struggled to articulate it through a prompt, Mei would still review AI's outputs to seek potential linguistic inspiration. As she noted: "Sometimes, in a whole paragraph it generates, there might be one word that stands out to me, and I think, this word is good, this word is useful, so I take that word and use it."

Table 1 shows an example of a paragraph from her first draft, her AI prompt, the AI's edits, and the final version. As a description of her book project, she felt the first draft of her paragraph was "quite plain," thus prompting AI to make it more "tempting." Mei evaluated the AI edits as "废话文学 [verbose writing]" and "purple prose" due to the unnecessary expanded length and "花里胡哨 [flowery]" wording, but acknowledged that they were "indeed more exciting" than her original version. While discarding most of AI's edits, she noted liking the phrase "shifting the lens from Y to the vibrant…" because, unlike her original wording, "contrary to viewing X as Y," the new phrasing suggests a more nuanced shift from the norm rather than a direct opposition. She also noted that in the argument sentence, the phrasing of the relationship between A, B, and C (underlined in Table 2) was refined for more clarity and conciseness. In her final edit of the argument sentence, she transferred the AI wording "vibrant" from the other sentence to replace "rich," rewrote the adverbial clause starting with "contrary to," and accepted AI's syntactic simplification. These changes allowed her to more effectively achieve her rhetorical purpose of making the writing more active and compelling. Mei regarded this process of critiquing, drawing inspiration from, and adapting AI's lexicon and syntax as a learning experience.

**Table 2**

*An Example Showing Mei's Pre- and Post-AI Edits*

| Mei's first draft | [Three sentences outlining her book's topic, goal, and methodology.] Contrary to viewing X as Y, the study reveals that it was the <u>crosscutting interactions between A and B, A and C, and among A within the system</u> that [the argument]. |
|---|---|
| Mei's AI prompt | Can you make this book description more tempting? |
| ChatGPT's edits | [Four sentences outlining her book's topic, goal, and methodology, the length of which is as twice as the original.] This groundbreaking book project offers a fresh perspective, shifting the lens from Y to the **vibrant** tapestry of human interactions within it. Prepare to be captivated as the narrative unfolds, revealing <u>the rich **interplay between A, B, and C**</u> that … [Two more sentences on the argument.] |
| Mei's final version | [Three sentences outlining her book's topic, goal, and methodology, without any change from the first draft.] Going beyond a conventional depiction of X as Y, the study shed light on <u>the **vibrant interplay between A, B, and C**</u> that [the argument]. *(Note: the wording was slightly changed to avoid online tracing.)* |

Design elements shaped Mei's perception and use of AI. Viewing GAI as "just a language model," Mei believed its greatest strength laid in its ability to "put things into language," leading her to treat it as a linguistic resource. She further explained why AI was effective for language-related tasks:

> It's not that I don't know a particular word, but I just can't recall it at the moment—I can't pull it from my brain's database. My thinking is more point-to-point, connecting a few things, while AI's thinking is more networked, drawing from a much larger database.

Because of AI's sophisticated language processing capabilities, Mei believed it could "represent the perspective of a reader," interpreting her writing in ways that might resemble a reader's understanding. Thus, she saw this as an opportunity to use AI to identify the ambiguity in her writing and improve clarity.

Furthermore, her metacognitive awareness of herself as a writer in relation to social factors led her to reflect on a fundamental ethical consideration regarding the use of AI in writing. Mei mentioned that while writing in her office, she kept ChatGPT open, but always minimized the window to prevent her colleagues and students from seeing it. She attributed her attempt to conceal her AI-assisted academic work to the "stigma" surrounding the use of AI in academic literacy practices, describing it as "using AI makes the work seem unoriginal or gives the impression of being less original." One particular experience that reinforced her sense of the "stigma" came from a pedagogy workshop at her school, during which her colleagues expressed "hostility" towards AI use in student writing. Mei defended herself and other students against what she perceived as the faculty's one-size-fits-all, extreme stance on AI: "We're not using ChatGPT to write the whole thing; we only consult it when we encounter ideas hard to express and go back and forth with it. But they [colleagues] were very resolute. They were firmly against using AI for editing or assisting with thinking." This experience left Mei feeling "concerned" and "afraid" to share her use of AI with colleagues, worrying if others would view her work differently.

Recognizing the conflict between her own ethical beliefs and those of others, Mei chose to keep her use of AI not only self-directed but also private. Over the span of the study, as she gained deeper metacognitive insight into her use of AI, her beliefs were strengthened, making her feel less "afraid" of external judgements:

> I've realized that I'm not as afraid as I used to be. On one hand, I know that some people will certainly think that way. On the other hand, I feel less insecure because I know how I use AI, and I know that my ideas are original. Unless someone could profoundly convince me otherwise—like if someone managed to convince me that my interaction with AI wasn't part of my thinking. They would have to make me believe that only ideas originating entirely from my own mind count as my thinking, and that my back-and-forth dialogue with AI doesn't count. Unless they can overturn my understanding, I believe this is justifiable.

However, Mei recognized that this external skepticism about work ethic and integrity, over time, also sparked other internalized concerns about her own abilities as a writer. She began to question whether using AI assistance discredited her writing ability and her capacity to "manage on my own," a quality she believed one should possess as an experienced academic writer. She used words such as "shortcomings," "weakness," "deficiency," "lack of fundamental ability" to describe her need to use AI for language suggestions. These words showed that being in a professional community largely critical of AI in writing, Mei started to see her use of AI as a reflection of her inadequacy as a writer, placing her competence in direct opposition to her resort to AI assistance.

Mei's primary goal in AI-assisted writing was to achieve clarity, accuracy, and linguistic richness, which she identified as an important quality in writing in her discipline. As she explained, "In our field, the primary quality is clarity. I want to express my ideas clearly and have others understand them as intended. I don't want readers to be confused about any meaning." As a humanities scholar, Mei was particularly mindful of how language, even down to a single word, could shape her writerly image and voice. Her attention to wordsmithing meant that she consistently sought language that best conveyed her intended meaning. She used her disciplinary and rhetorical knowledge to guide her judgment in applying AI, as she exemplified: "highly formal and official language can be appropriate at times, but sometimes it feels out of place. So the judgment of whether to use a word such as 'embark on' or 'start' is based on having a clear understanding of what I want." The process of recursive communication with GAI allowed her to mobilize and orchestrate the linguistic resources from both herself and the LLM, collaborating in the ongoing and emergent refinement of meaning. Mei acknowledged that the clarity AI helped her achieve enabled her to better "present my voice" in her academic writing.

### 4.2 Daniel

Daniel's entering motivation to use AI came from his concern about being "not capable of" addressing certain writing tasks. He used GAI extensively throughout his writing processes, such as refining his research scope, organizing and presenting results, interpreting data, and writing academic reports. He described his writing process with ChatGPT as follows:

Previously, I would write a sentence or a paragraph, think it over, and revise it to make sure it was appropriate. Now I just throw everything I want to say at GPT without worrying about grammar. I let it handle the grammar and then review the output. If it conveys my intended meaning, I copy and paste it. If the revision isn't exactly what I want, I make further adjustments based on what GPT provided. Then if the changes are what I want, I use it directly; if not, I keep modifying it.

As this quote shows, in both non-AI and AI-assisted writing, Daniel employed various cognitive strategies such as monitoring, evaluating, and adapting. Yet, by relinquishing some of his cognitive effort to AI, Daniel shifted from being the architect and builder of the text who was in charge of drafting and crafting every element, to also assuming a role of mediating AI. In this role, AI acted as a proxy to assist in constructing the building blocks of his text, while he monitored, adapted, and directed the AI's outputs to fit into his overall vision. Daniel actively engaged in this reorientation of cognitive strategies to his interaction with AI, where he constantly negotiated meaning. This is evident in his prompting process: When Daniel identified inaccuracies in ChatGPT's outputs, he pointed them out: "No! I think it should be A, why do you think it is B?" Alternatively, he would explain what he believed the correct answer should be: "No. X means [explanation] and Y means [explanation]. If it is Y, then it should include [A element] as shown in the definition." When the AI insisted on its answer, Daniel uploaded a relevant source as evidence, prompting the AI to reconsider: "Look, this is what it says." Daniel noted that AI would acknowledge its mistake when it realized it was wrong; if the AI maintained its stance, it would explain its reasoning in greater detail, which he would then carefully review to consider its reasoning. As he stated, "This process allows me to engage deeply with GPT, evaluating its outputs, my reasoning, and external sources."

In this process, Daniel treated AI less like a mechanical system that passively carries out tasks but more like an active participant by putting significant effort into the process he described as "teaching AI." By questioning, presenting counterarguments, and referencing external sources, Daniel prompted the AI to reconsider its stance, engaging in a process that mirrored teaching human thinking. Daniel's approach was not teacher-centered but Socratic–an iterative and dialogic dynamic that went beyond simply correcting student mistakes and showing the right answers. His openness to understanding the reasoning behind AI's responses also fostered an interaction where the human and AI collaboratively learn, adapt, and evaluate different perspectives.

Daniel's use of AI was shaped by various design, support, and social elements. His beliefs in ChatGPT's capabilities were based on his understanding of AI's design. As he noted, "because its training data has a default understanding of the topic," it needs specific contextual information to override its default, generic answers. Thus he always provided contextually rich prompts to anchor AI's analytical abilities in the specific topic he was writing about. As he stated: "If you ask GPT to find literature for you, it's very unreliable. But if you feed it the materials and ask it to dig deeper, it can be more reliable. You need to feed the content first, and then it can effectively analyze and process the information."

Apart from building connections between AI and other nonhuman resources such as academic articles as a distinct way of managing his learning, peer and teacher support was also a crucial support element in helping him leverage AI resources. Over the span of the study, Daniel expanded his use of AI from only assisting with writing to utilizing it across all stages of his research project. He recalled the first time he used AI to locate suitable theoretical frameworks, as he lacked the disciplinary knowledge needed to conduct research for a course outside his field:

I didn't initially think of using ChatGPT to find the theoretical framework because I wasn't familiar with X [field name] theories. I had no idea where to start. Then, a friend of mine, who's also a graduate student at my school, suggested that I feed all my results to GPT and see if it could help. So I tried it, and it turned out to be quite useful.

From the list of options provided by AI, Daniel selected two and continued exploring them by interacting with AI and searching for additional academic sources in which the theory was used. He then scheduled a meeting with the professor to finalize his choice. Reflecting on his learning, he said, "I didn't initially understand the framework, but after going through this process, I finally understand what the theoretical framework is used for."

Even though Daniel claimed that he learnt a lot because of his engagement with AI, it was not without ethical concerns. He admitted that despite the professor's "no ChatGPT allowed" policy, he used it without "feeling guilty," as he believed that he did not use it "to do things I shouldn't be doing," which, to him, meant copying others' work and creating unoriginal products: "Since the content is originally mine and I only use AI for revisions, I don't consider it plagiarism. I think using AI to make my writing clearer actually benefits the reader, making it easier for them to understand." He viewed the professor's policy as "old-fashioned" and further justified:

> In today's society, we can't go back to the way things were as the professors wanted. I
> believe that instead of banning AI, teachers should teach students how to use it to
> enhance their efficiency because we live in an era with an overwhelming amount of
> information. Even if students copy directly, it should be the teacher's responsibility. It's
> the teacher's responsibility to guide the students. The concept of authorship has become
> quite blurred in this age, and it's important to adapt to this change.

Daniel's employment of AI resources also depends on his evaluation of the nature of the task and his purpose of learning in a specific context. He differentiates the depth of his critical engagement with AI, contingent on the significance he personally attributes to a task: "If the task is not high-priority, like writing an annotated bibliography, I'll usually glance over the changes GPT made, and if everything looks fine, I'll submit it. It saves a lot of time." For tasks he deemed more important, Daniel engaged in a highly patient, dialogical inquiry with AI, meticulously working with it to negotiate and co-construct meaning. This differentiated approach reflects his view of human cognition and AI not as separate or hierarchical, but collaborative and interrelated. In other words, AI was not treated as an external tool for executing the human will; rather, it became an integral part of both his language repertoire that was orchestratable into meaning-making, and at the same time, a medium and a process for facilitating and interrogating such meaning-making processes.

### 4.3 Alex

Despite the wide range of AI-assisted writing tasks Alex engaged in across his academic work, language "polishing" emerged as his primary motivation for using ChatGPT—a theme consistently highlighted both in the interview and his AI prompts. Prompts like "Is there any grammatical issues with the sentences?," "Polish the following sentences," "Polish and enrich the following paragraph," "Enhance and polish the following sentences," and "Polish the following review comments" appeared repeatedly in his AI chat history. Alex used ChatGPT to polish his research papers, professional emails, responses to reviewers, review comments he gave to others, and his social media posts promoting his work. This clear, yet broad goal setting characterized Alex's overall use of AI. As he exemplified:

I write a passage myself and give it a command, like asking it to polish these sentences. Then it will generate a polished version of the sentences, and I'll review it from start to finish to see if there are any issues, mainly logical issues. If there are, I'll make adjustments. If not, I'll use the polished sentences as my final version.

Alex's use of the word "command" reflects his perception of AI as a passive tool that operates, and perhaps should always operate, under the user's control. His AI logs further revealed that much like his goal setting, his approach to monitoring, evaluating, and adapting AI usage also followed a consistent pattern–generic and linear in nature. When dissatisfied with the AI's edits, he often re-entered the AI-generated version back into the system, using the same generic prompt—"enhance and polish the following sentences"—repeatedly until it achieved a satisfactory result. Occasionally, Alex would adjust his prompts to include a little more contextual or content-specific information, such as: "Enhance and polish the following sentences to explain how to select important variables based on X model." Even for tasks that might require a different tone or register, such as emails or social media posts, he relied on the same prompts.

Alex's ethos of using AI was shaped, in part, by his (lack of) understanding of the technology's design elements. As he said: "Although I use it frequently, I think my understanding of it is still just on the surface level. I don't have much knowledge of its internal logic or how it generates things—those might be more related to computer science." However, what perhaps more deeply influenced his use of AI was his discipline-based view of what constitutes good writing, especially for L2 writers. As he stated:

I feel that for writers using English as a second language, the most important use of

ChatGPT is still article polishing, to avoid grammatical errors. This is because grammatical accuracy is crucial. In [his field] writing, as long as the meaning is clearly and concisely expressed, even using short sentences is sufficient. This differs from other types of writing which requires a more complex style. Therefore, the main fear for me as a non-native writer is making grammatical mistakes, which is why I primarily use ChatGPT for this purpose.

Alex's pursuit of grammatical accuracy was rooted in his belief that "native-speaker level fluency" is the golden standard for writing: "My writing definitely had that Chinglish feeling during my first year or two studying in the US. In recent years, it's getting a bit better, especially after ChatGPT came along, my writing feels more like a native speaker." Alex emphasized the emotional benefits as a result of using AI, describing it as providing a sense of "security," making the writing process "less of a struggle," and "not the headache it used to be." This extra scrutiny and awareness of language mechanics, along with a preconceived notion about L2 writing, also surfaced in his academic review work of other L2 writers. He admitted that even in double-blind reviews, he could immediately tell if the author was Chinese based on the language, thus impacting how he approached the review:

When I knew the author was Chinese at first glance, I would immediately wonder whether their writing or expression had any grammatical errors. I would pay extra attention [to language]. And if the writing isn't very smooth, the first feedback I usually give is to suggest proofreading the whole paper.

Alex regarded his use of AI "without ethical concerns." When submitting his review comments, he responded "no" to journal systems' inquiry about whether GAI tools had been used to generate feedback. In Alex's view, using AI for "language polishing" did not constitute "using AI to generate feedback," thus he saw no need to disclose his AI usage. As an experienced writer and reviewer, Alex navigated these two perspectives fluidly, taking into consideration how using AI in his writing could influence reviewers' judgment. Ultimately, in Alex's practice, GAI served to eradicate grammatical errors and neutralize written accents, allowing his writing to more closely approximate that of a native speaker, which he perceived as desired in his discipline. This alleviated much of the anxiety he previously experienced with L2 academic writing, enabling him to conform to the disciplinary norms of language accuracy that he valued.

## 5. Discussion

The three cases show a spectrum of self-directed AI use in academic writing, all related to the participants' concerns as L2 writers. While they actively monitored their AI-assisted writing process—planning, evaluating, and adapting the use of AI—their approaches varied significantly due to their differing understandings of AI's design and their positionality in relation to AI. One notable distinction is the *prescriptive* versus *dialogic* uses of AI as a language resource. A prescriptive approach, as shown in Alex's case, mainly treats AI as an editing tool that follows the user's command to perform linguistic tasks, identify issues, and fix the text. This positionality assumes the writer's full control over the text, rejecting AI's agency and its potential discursive role in participating in meaning construction. In this way, Alex positioned AI as a mechanical, closed system of words governed by grammatical rules and algorithms, allowing it to mechanize and standardize an L2 writer's text according to pre-set rules that perfect grammar. While Alex perceived himself as retaining full control over the text by denying AI's agentic role, this approach dehumanizes writing by pushing it toward AI-driven standardization. In contrast, Mei and Daniel demonstrated a more dialogic use of AI, positioning it as a heuristic repertoire. Mei, for instance, treated AI as an interactive and interpretive audience–a "social body" (Anderson, 2023) materialized in words to collaboratively interrogate the clarity of her writing. By positioning it as a social entity rather than an inanimate tool, her recognition of AI's agency also makes her writing process more socially constructed. Similarly,

Daniel's heuristic process with AI to negotiate meaning shows a more posthuman orientation in which AI embodies part of Daniel's distributed cognition that builds dialogue between his various linguistic and thinking resources, human and nonhuman. As Pennycook (2018) suggested, this orientation "step[s] out of the humanist constructs of the individual and the community and look instead at the notion of distributed language and spatial repertoires" (p. 445). Daniel's dialogic approach positions AI not as a tool or a closed system, but as fluid semiotic resources in a state of *becoming* – actively participating in Daniel's meaning-making – instead of *being* static words, rules, and systems. Daniel de-centers the human as the only rhetorical agent that dictates meaning and treats meaning as emergent between human-AI intra-actions (Author, under view). The distributed cognition and agency unsettle anthropocentric boundaries in semiosis, embracing the fluidity of human-AI assemblage as a contribution to his academic literacy practices.

The participants' differing positionalities toward AI are not by chance, aligned with the perceived role of language and literacy in their academic work, grounded in their writerly *habitus* (Bourdieu, 1971) developed through their L2 academic writing experiences within the contextual relationships of their disciplines. Valuing the author's distinct voice in humanities writing, Mei meticulously scrutinized AI's language inputs and carefully attended to its diction and register. Mei's approach also resonates with the humanities' interpretive tradition towards meaning and text as not merely descriptive but also constructive. STEM scholar Alex treated both AI and text as a tool/vehicle that carries predetermined meanings by the writer. His focus on using AI to achieve linguistic accuracy and correctness assumes a reification of a functionalist perspective toward language as a carrier of scientific reasoning, which is best transmitted through standardized linguistic forms. Accented L2 writing and deviations from standardized expressions can introduce ambiguities and confusion that undermine scientific understanding. While Alex's use of AI can be proper given his understanding of the rhetorical context within his discipline, it raises important questions about its broader applicability for other L2 writers. While one perspective may see this use of AI as "promoting equity" for L2 writers (Giglio & Costa, 2023), it risks perpetuating native speakerism and reinforcing standardized language ideologies that have long dominated academic writing and publishing, further marginalizing non-Western rhetorics (Author, under review).

Regardless of a prescriptive or a dialogic approach to AI in writing, one cannot deny the intertextual nature of what emerges from the interaction between human and AI, prompting important reconsiderations of the relationship between text and authorship. The notion of a singular, authoritative author was under question when the self-directed users adapted AI inputs through iterative cycles – not only is it hard to trace which part of the text is human or machine, but also the AI input itself is untraceable, as it was sourced from countless other human texts. For over five decades, poststructuralist thinkers such as Barthes, Foucault, and Derrida have been challenging authorialism under the structural paradigm. Barthes (1967) critiqued authorship as the "epitome and culmination of capitalist ideology" (p. 142). Foucault (1984) viewed texts as discursive constructs, and the author as a function of discourse and power–rather than the originator of meaning. Derrida's (1978) concept of différance also undermines the author as a sovereign owner, empowering the text to be multivocal. Différance, connoting both notions of difference and deferral, unveils how meaning is endlessly deferred and constructed rather than present and given. A poststructuralist orientation positions texts – human or AI – in a constant state of *becoming*. The meaning emerging from human-AI inter-actions is also constantly deferred and in flux as L2 writers engage in decontextualizing and recontextualizing AI outputs in their writing process. In this notion, authorship is less about claiming a definitive origin and controlling a final text than about recognizing author-*ing* as an evolving process of entextualization, where writers' spatial repertoires –both human and nonhuman–are intricately assembled through the interplay of presence and absence, enmeshed in a broader network of relationality and a dialectical state of being and becoming. While this is still a debatable issue, we offer one perspective that considers how L2 writers' engagement with AI may challenge traditional notions of authorship.

What is perhaps further disrupted, as shown in the three cases, is the relationship between learners, learning, and learning materials. GAI offered unique affordances for participants to self-direct their learning about writing, enabling them to selectively and strategically engage with different components of the writing process and define the nature of learning entailed. As seen in Mei's and Daniel's cases, SDL with AI demands strong motivation to self-monitor and manage one's learning, along with deep cognitive and metacognitive engagement that involves critical and reflective thinking about AI's role in the process. Yet Alex's use of AI may resonate with common concerns, such as that AI may be used to "avoid education" (Gefen et al., 2023) and that "the foundational writing skills required to use AI effectively and the actual use of AI-assisted writing can be in contradiction to each other" (Warschauer et al., 2023). These concerns frame writing development and AI skills as oppositional, if not at least separate, creating an unproductive division given the increasingly self-directed use of AI in writing and language learning. We thus highlight the importance of considering AI use as part of (critical) rhetorical and digital literacy (Warschauer et al., 2023; Darvin, 2023; Gupta et al. 2024; Author, 2024), so that emergent forms of SDL can be leveraged for literacy development rather than being dismissed. Both Mei and Daniel exemplified this: they ultimately developed strong convictions about the value of their interaction with AI as intellectually worthwhile. Mei developed a clearer sense of her own ethics, viewing AI as a legitimate part of her intellectual work rather than a form of plagiarism. Daniel, on the other hand, questioned his professor's policy, arguing that it no longer adapted to the evolving concept of authorship today. However, challenges arise to render this learning visible and accessible, as shown in the significant and nuanced differences in how the three writers used AI.

Stanton (2023) noted that many academics focus on preventing plagiarism and limiting AI, but this overlooks the more critical need for developing new ethical frameworks. At the heart of why using AI can be easily regarded as problematic and stigmatized lies the conventional understanding of the textual-authorial relationship under which the learner-writer identity is defined. The participants in the study kept their AI usage confidential, even when disclosure was required, which shows the failure of the current AI policies to address such ethical complexities. One reason for withholding the information is their fear of ambiguous or negative connotations and judgment associated with AI use in academic writing, fostered by a larger discourse that frames AI as the chief culprit in a new literacy crisis. In this sense, AI becomes a social shaming technology that devalues L2 writers' SDL, which risks reinforcing the long-standing deficit view on L2 writers' language abilities and their credibility. Hence beyond building the myriads of ethics around AI, morally and practically (for instance, see O'Regan & Ferri, 2024) for the social goods, we should also be cautious of the consequences of ethical judgments when an L2 learner confronts the evolving social anchoring of ethics that possibly devalues their ethos. However, we do recognize that some of the participants' usage of AI can raise critical ethical concerns that necessitate further discussion (e.g., feeding AI with copyrighted materials).

**6. Conclusion**

This study explores the self-directed use of AI and the learning possibilities it presents, particularly within the underexplored context of advanced L2 writers' scholarly writing practices. The findings challenge two prevailing assumptions about GAI in learning: first, that GAI allows one to bypass learning and cognitive engagement by producing seemingly correct answers; second, that while GAI may support undergraduate learners, it is not suitable for advanced learners (Gefen et al., 2023). Contrary to these views, we found that advanced, experienced L2 writers deeply engaged with GAI to seek meaningful learning and to address their L2-specific writing challenges. We found that their approaches (e.g., prescriptive vs dialogic) and positionalities (e.g., tool vs participant) varied significantly, showing the potential of engaging with AI to transform academic writing and influence existing inequalities for L2 writers both in productive and disruptive ways. These findings show how the intertextuality involved in students' SDL process of writing with AI disrupts traditional notions of authorship, text, writer, learner, and learning. We advocate for new ways of understanding these disruptions through a poststructuralist view on language and writing practices beyond the human-AI, writing-technology, and learning-bypassing divides. Moving beyond the current counterproductive discourses that reinforce these dichotomous thinking is imperative, as they risk further disadvantaging L2 writers. We advocate for collectively deconstructing and reconstructing the social body of AI, to engage with it in ways that foster writerly virtues – through their voice, autonomy, confidence, and positioning (as cited in Daniel et al., 2023) – in the postdigital age. GAI places us at a pivotal moment to address these long-standing challenges in our education and our historical relationship with technologies.

Lastly, we acknowledge that due to our small sampling size, the findings may not capture the complex and nuanced ways of using AI technologies for academic purposes by advanced L2 writers. Moreover, since all three participants share the same L1 and a similar educational background, their practices of using AI in writing may not represent L2 writers of different linguistic and cultural backgrounds. Thus, we want to further question whether SDL practices with AI could be influenced by the larger social and cultural context in which AI tools are commercialized and regulated. Future studies could look into the impact of public discourses on AI tools on how L2 writers position and employ AI tools for a wider variety of purposes, both within and beyond the educational context.